\magnification=\magstep1
\hsize 32 pc
\vsize 42 pc
\baselineskip = 24 true pt

\centerline {\bf A Quantum Mechanical Derivation of Gamow's Relation }
\centerline  {\bf For the Time and Temperature of the Expanding Universe}

\vskip .3 true cm
\centerline {\bf Subodha Mishra and D. N. Tripathy$^\dagger$}
\centerline{ \it Department of Physics and Meteorology \& Centre for Theoretical Studies}
\centerline{ \it Indian Institute of Technology, Kharagpur-721302, India}
\centerline { \it Institute of Physics, Sachivalaya Marg, Bhubaneswar-751005, India$^\dagger$}
\vskip .4 true cm
\vskip 2 true cm
\centerline {\bf Abstract}
\par The quantum mechanical approach developed by us recently for the evolution of the universe is used to derive
an alternative derivation connecting the temperature of
the cosmic background radiation and the age of the universe which
is found to be similar to the one obtained by Gamow long back. By assuming the age of
the universe to be $\approx$ 20 billion years, we reproduce a value
of $\approx$ 2.91 K
for the cosmic back-ground radiation, agreeing well with the recently
measured 
experimental value of 2.728 K. Besides, this theory enables us to calculate the photon
density and entropy associated with the background radiation and the
ratio of the number of photons to the number of nucleons, which
quantitatively agree with the results obtained by others.

\vfill
\eject

It has been, by now, accepted that the most important theory for the
origin of the universe is the Big Bang Theory [1], according to which
the present universe is considered to have started with a huge 
explosion from a superhot and a superdense stage. Theoretically
one may visualize its starting from a mathematical singularity with
infinite density. This also comes from the solutions of the type I and type II
form of Einstein's field equations [2]. What follows from all these solutions is 
that the universe has originated from a point where the scale factor $R$ (to be 
identified as the radius of the universe) is zero at time $t = 0$,
and its derivative with time is taken to be infinite at this time. That is, it is 
thought that the initial explosion had happened with infinite velocity,
although, it is impossible for us to picture the initial moment of 
the creation of the universe.

An indication in support of the Big Bang theory is the expansion of the 
universe, which has been established by means of the Hubble's law [3] as
$$ v = H_0d  \ , \eqno{(1)}$$
where $v$ is the radial recession velocity of the galaxy and $d$ is the 
distance of the galaxy from us and $H_0$ is known as the Hubble constant. The
quantity $(1/H_0)$, which is known as the Hubble time, is a measure for the
maximum age of the universe. Since, it is very difficult to correctly
determine the distance $d$ to the galaxies, there is a great uncertainty in the
estimated value of $H_0$. The correct value for the age of the
universe seems to lie in between 10 to 20 billion years. The most
important evidence for the Big Bang theory is the microwave background
radiation, which was discovered by Penzias and Wilson [4] with an effective
temperature of $\approx 3.5 K$. However most recent measurement on
the cosmic background radiation using Far Infrared Absolute
Spectrometer (FIRAS) [5] has yielded a value of 2.728 K. 
The characteristic of this radiation is that
it is almost absolutely isotropic, that is, it comes to us from all directions
with the same intensity. This means that the radiation is not due to stars or galaxies which
are the measure of the inhomogeneities of the universe. The only plausible
explanation for the origin of the cosmic background radiation is
that the universe has, perhaps,
passed through a state of very high density and high temperature in its early
state, and the present temperature of $\approx 2.728 K$ [5] is nothing but the remnant of the
intense heat of the Big Bang, which has been redshifted into the microwave region.

Since, within a time period very close to the Big Bang explosion $(t
= 0)$, the universe was
lying in its `radiation dominated era', there was no possibility for the 
formation of elementary particles with finite mass  at that stage. The actual
creation of material particles must have taken place a few seconds after the 
Big Bang. In the mid 1970s, Gamow [6] suggested that the high density and high
temperature required for the synthesis of elements, existed in the few moments after 
the Big Bang. In his simplified picture, Gamow assumed the universe to be 
intially made of neutrons and photons. As one knows, the neutrons are charge free
particles found in the nuclei of atoms, while photons are the quanta of the 
electromagnetic field that constitute light. Gamow arrived at a relation, relating
the temperature $T$ of the universe with time $t$ after its Big Bang, which is given
as
$$ T = \bigg ( {3 c^2\over 32 \pi G a}\bigg )^{1/4} t^{-1/2} K  \ , \eqno{(2a)}$$
where $a$ is the radiation constant and all other constants in the above
equation have their usual meaning. A numerical estimate of the factor within
the bracket in the above equation gives
$$T = 1.5 \times 10^{10} \ t^{-1/2} K  \ . \eqno{(2b)}$$
Later on, the above relation was modified by Hayashi [7] by taking into account 
the effects of the thermal equilibrium among the particles, like neutrons, protons,
electron-positron pairs and neutrino-antineutrino pairs, present in the universe at
the very temperatures that existed right at the beginning of the universe.
Hayashi obtained a relation, which is given as
$$ T = 10^{10} t^{-1/2} K  \ . \eqno{(3)}$$
As one can see from Eq. (3), it differs from Gamow's derivation
[Eq. (2b)] with respect to the extra factor of (3/2). Based 
on the relation $(2b)$, Gamow made the prediction that a very faint back-ground of
radiation, known as the relic of the Big Bang, should exist at the present
epoch of the universe. This was subsequently verified by Penzias and Wilson [4], who 
reported an isotropic radiation background with a temperature of
$\approx 3.5 K$, in
the microwave region. If one takes Eq. (3) to be correct, then, to reproduce a
temperature of $\approx 2.728 K$ the present age of the universe would be $\approx
425$ billion years instead of 20 billion years, where the latter one has been known to be very close
to the accepted value. Gamow's relation would give a value of $\approx 956$ 
billion years for the age, which is much more absurd.

Recently, we have developed a quantum mechanical theory [8] for a system of self-
gravitating particles like the stars that have exhusted all the
nuclear power at their respective cores. In these systems, the particles
interact with each other gravitationally. Using a
singular form of single particle density to account for the distribution of
particles within the system, we have been able to obtain a compact expression for the
radius of a neutron star. Comparing
this with the Schwarzschild radius, we arrive at a critical value [8] for the
mass of the neutron star beyond which it should go over to the stage of a 
blackhole. Our value for the critical mass, seems to agree with those of other theoretical calculations.
Applying such a theory to a white dwarf, we have succeeded to reproduce the
so called Chandrasekhar limit [9] for its critical mass. In a
subsequent work, we have used the above
theory [10] to make a study about on the evolution of the present universe by visualizing
it as a system constituted of a large number of self-gravitating ficticious
particles, fermionic in nature, interacting with each other through gravitational
potentials. As far as the neutron stars are concerned, they are
obviously constituted of fermions. For the universe, it is being said
that the major constituent of the total mass of the present universe
is made of the Dark Matter (DM). Since neutrinos are considered to be
the most probable candidates for the particles of the DM, we are
justified to say that the universe is constituted of particles that
are fermions.
Proceeding in a manner similar to the one used by us [8] for the
study of stars,
we arrive at an expression [10] for the radius of the universe which, after invoking
the Mach principle [11], assumes a form involving only the fundamental constants
like $G$, $\hbar$ , $c$ and the mass $m$ of the constituent particles. Following
this  expression, we have made an estimate of the total mass of the universe, which
is found to be agreeing with the results of other theoretical calculations [12]. 
Our calculated value for $(\dot G/G)$ is also in good agreement with those of many earlier workers.
There are many other interesting results that follow from this theory, which
demand to have a deeper study on the subject. In this present paper, we want to apply
the very theory [10] to make an estimate of the temperature of the cosmic background
radiation, whose most recent
value has been reported to be $\approx 2.728 K$ [5]. Using our expression, we also
try to discuss about the production of the various elementary particles that took place in
the early universe. All these have been dealt with, in the next section.

The Hamiltonian used by us recently [8,10] for the study of a system of
self-gravitating particles is written as  
$$H = - \sum^N_{i=1}({\hbar^2\over 2m})\nabla_i^2+{1\over 2}\sum^N_{i=1}
\sum^N_{j=1,i\not =j} v (\mid\vec X_i-\vec X_j\mid) \ , \eqno{(4)}$$
where $v(\mid\vec X_i-\vec X_j\mid) = - g^2/\mid\vec X_i-\vec
X_j\mid$, having $g^2=Gm^2$, 
$G$ being the universal gravitational constant and $m$ the mass of the
constituent particles whose number is $N$. 
Since the measured value for the temperature of the cosmic microwave
background radiation is $\approx 2.728 K$, it lies in the neighbourhood
of almost zero temperature. We,therefore, 
use the zero temperature formalism for the study of the present problem.
Under the situation $N$ is
extremely large, the total kinetic energy of the system is obtained
as 
$$<KE> = \bigg ({3\hbar^2\over 10m}\bigg )(3 \pi^2)^{2/3}\int d\vec X
[\rho(\vec X)]^{5/3} \ , \eqno{(5a)}$$
where $\rho(\vec X)$ denotes the single particle density to account
for the distribution of particles (fermions) within the system, which is
considered to be a finite one. Eq. (5a) has been written in the
Thomas-Fermi approximation. The total potential energy of the system,
in the Hartree-approximation, is now given as 
$$<PE> = - ({g^2\over 2})\int d\vec X d\vec X'{1\over \mid\vec
X-\vec X' \mid} \rho(\vec X)\rho(\vec X') \ . \eqno{(5b)}$$
Inorder to evaluate the integral in Eq. (5a) and Eq. (5b), we had
chosen a trial single-particle density $\rho(\vec X)$ [8,10] which was of
the form :
$$\rho(\vec X) = {{A e^{-x}}\over x^3} \ , \eqno{(6)}$$
where $x = (r/\lambda)^{1/2}$, $\lambda$ being the variational
parameter. As one can see from Eq. (6), $\rho(\vec X)$ is singular at
the origin. Its interpretation has already been given in our
earlier papers [8,10]. After evaluating the integrals in Eq. (5) to
find the total energy
$E(\lambda)$ of the system, we minimize it with respect to $\lambda$
and thereby we obtain the value of the energy of the system
corresponding to its lowest
energy state (ground state). Following the expression for $<KE>$
evaluated at $\lambda = \lambda_{min} = \lambda_0$, we write down
the value of the equivalent temperature $T$ of the system, using the
relation 

$$\eqalign{ T = & ({2\over 3})({1\over k_{\beta}}) [ {< KE >\over N}]\cr
= &  ({2\over 3}) ({1\over k_{\beta}}) (0.015442) N^{4/3} ({mg^4\over
\hbar ^2})\cr}  \eqno{(7)}$$

The expression for the radius $R_0$ of the universe, as found by us
earlier [10], is given as
$$R_0 = 4.047528 ({\hbar^2\over mg^2})/N^{1/3} \ . \eqno{(8)}$$
After invoking Mach's principle [11], which is expressed through the
relation $({GM\over R_0 c^2})\approx 1$,  and using the fact that the
total mass of the universe
$M= Nm$, we are able to obtain the total number of particles $N$ constituting the
universe, as 
$$N = 2.8535954 ({\hbar c\over Gm^2})^{3/2} \ . \eqno{(9)}$$
Now, substituting Eq. (9) in Eq. (8), we arrive at the expression for
$R_0$, as
$$R_0 = 2.8535954 ({\hbar\over mc})({\hbar c\over Gm^2})^{1/2} \ . \eqno{(10)}$$
As one can see from above, $R_0$ is of a form which involves only the fundamental constants like $\hbar, c, G$ and
$m$. Now, eliminating $N$ from Eq. (7), by virtue of Eq. (9),we have
$$T = {2\over 3} (0.0625019)({mc^2\over k_{\beta}}) \ . \eqno{(11)}$$

Let us now assume that the radius $R_0$ of the universe is
approximately given by the relation
$$R_0 \simeq ct \ , \eqno{(12)}$$
where $t$ denotes the age of the universe at any instant of time. 
The Hubble's law as indicated in Eq. (1), also implies that the universe
is expanding uniformly. Although, it is so for the universe, all the
galaxies are not uniformly expanding. Considering a photon of light with wave length
$\lambda$ travelling a distance of separation 'd' between two
galaxies at rest with respect to
each other, one has $d=ct$, where 't' is the time it takes for light to
travell the space between the galaxies. Because of the expansion of
the universe, the galaxies move away from each other at a velocity
$v$ known as the radial velocity. During this time 't', the galaxies
are separated by a distance $\Delta d$ given by $\Delta d = vt$.
Thus, one finds that ${\Delta d \over d} =({v\over c})$. From this, it follows that the
greater is the relative velocities of the galaxies, the greater is
the separation attained in the time interval $t$. The
importance of Hubble law, as stated through Eq. (1), is that the galaxies were closer in the past
than they are now. As we have stated earlier, the Hubble time $({1\over H_0})$ represents the
maximum age of the universe, because the galaxies themselves slow
down the expansion of the universe. Even though the galaxies are
farther apart, they still exert a gravitational force on each other.
Their mutual gravity continuously acts to pull other galaxies together. This
means that the universe was expanding faster in the past than it is
now. As indicated in Eq. (12), the velocity of expansion of the universe is being approximated
to be equal to the velocity of light 'c'.

Following
Eq. (10) and Eq. (12), we write $m$ as 
$$ m = ({\hbar^3\over Gc^3})^{1/4} (2.8535954)^{1/2} {1\over\sqrt t}
\ . \eqno{(13)}$$
A substitution of $m$, from the above equation, in Eq. (11), enables us to write

$$ T = 0.070388 ({1\over k_{\beta}}) ({c^5\hbar^3\over G})^{1/4} t^{-1/2}\eqno{(14a)}$$
$$ \ \  = 0.070388 [({c^3\over G}) {\pi^2\over 60\sigma}]^{1/4}
t^{-1/2} \ , \eqno{(14b)}$$                      
where $\sigma = ({\pi^2 k^4_{\beta}\over 60\hbar^3 c^2})$, is the Stefan-Boltzman constant [13]. 
Substituting the numerical value of $\sigma$, which is equal to $5.669\times 10^{-5} \
erg/cm^2. deg^4. sec$, and the present value for the universal gravitational constant
$G$ $[G = 6.67 \times 10^{-8} dyn. cm^2. gm^{-2}]$, in Eq. (14b),we obtain
$$ T = (0.23172 \times 10^{10}) t^{-1/2} K \ . \eqno{(15)}$$
As one can very well see, Eq. (15) is of the same form as obtained by
Gamow [Eq. (2b)] leaving aside the
multiplying constant 0.23172. If we accept the age of the universe to
be close to $20\times 10^9 yr$, which we have used here, with the
help of Eq.
(15), we arrive at a value for the cosmic background temperature
equal to $\approx 2.91 K$. This is very close to the measured value
of 2.728 K as reported from the most recent Cosmic Background
Explorer (COBE) satellite measurements [5]. However, to reproduce the
exact value of 2.728 K for the cosmic background temperature from our
expression, Eq. (15), we would require an age of $22.832 \times 10^9
yr$ for the universe.

By virtue of the expression given in Eq. (14b), we find
$$\sigma T^4 \simeq 2.4547 \times 10^{-5} ({\pi^2 c^3\over 60 G})
{1\over t^2} \ . \eqno{(16)}$$
The very form of the above equation suggests that the factor in its right
hand side (rhs) can be identified as the energy density of the electromagnetic
radiation at a time $t$. The radiation of this form is belived to follow the black-
body law.  The very agreement of our calculated
result with the most accurate value for the temperature of the
background radiation shows that age of the universe is very close to
$\approx 20 \times 10^9 yr$. This also creates a kind of
confidence in us regarding
the correctness of our theory compared to others, inspite of its
basic difference from the conventional
approaches, relating to the evolution of the universe.

Using Eq. (15), we have made an estimation of the temperature of the universe at
various stages of its evolution in time. Comparing the energy,
associated with the temperature $T$, with $mc^2$, we calculate the
masses of the elementary particles 
formed at various times. This is shown in table-I of this paper.
From the table, we notice that when the age of the universe
was less than 5 sec, the formation of electron and positron was possible,
while when the age of the universe was less than $1.2 \times 10^{-4}$ sec, the
formation of muons and their antiparticles must have taken place. For
the formation of mesons and their antiparticles, which needs a
temperature of $\approx 1.6\times 10^{12}K$, the corresponding age of
the universe would be less than $\approx 7\times 10^{-5} sec$ .As far
as the nucleon (neutron and proton) and their antiparticles were concerned, they must
have been formed before an age of $1.5 \times 10^{-6}$ sec. Thus, the period between
$t = 7 \times 10^{-5}$ sec and 5 sec may be called as the lepton era, while the period
before $7\times 10^{-5}$ sec is called the hadron era. The very early era
which is known as the planck era corresponds to the period $t < 10
^{-43}$ sec, $(temperature < 10^{32}K)$.
During this period, gravity is considered to be playing a major role
and it is to be, possibly, quantized at that stage.

Having evaluated the expression in the rhs of Eq. (16), the energy of the electromagnetic
radiation radiated per unit area per unit time is given as 
$$ u = 1.6345 \times 10^{33}({1\over t^2}) \ , \eqno{(17)}$$
where $t$ is the age of the universe in sec at any instant of time. The
entropy $S$ associated with the microwave back-ground radiation is obtained
as [14]
$$ S = {16 Vu\over 3 cT} = 2.9058 ({V\over T}) \times
10^{23}({1\over t^2}) \ . \eqno{(18)}$$
Assuming the present universe to be spherical, its volume $V$ is
given as  $V = ({4\pi\over 3}) R^3_0$, where $R_0$ 
denotes its radius. Taking $R_0\simeq 2.16 \times 10^{28}$ cm, which
corresponds to the age $t = 22.832 \times 10^9 yr$, since $(R_0 \approx ct)$, the 
photonic entropy of the present universe is calculated to give
$$S = 2.369 \times 10^{73} ({1\over T}) erg/deg  \ , \eqno{(19a)}$$
For $T=2.728 K$, it becomes,
$$S = 0.86\times 10^{73} \ \approx 10^{73} erg/deg \ . \eqno{(19b)}$$

The equilibrium number of photons [14] associated with the microwave background radiation
is given as
$$\overline N_{\gamma} = {V 2\zeta (3)\over\pi^2\hbar^3 c^3}
k^3_{\beta} T^3_0 \simeq (410.0) V \ . \eqno{(20)}$$
Following this, the photon density is found to be $({\overline N_{\gamma}\over V}) \simeq 410$, which
is in very good agreement with the estimated value of 400 found [15] by
doing a calculation of the total energy density carried by the cosmic
microwave background radiation.
Using Eq. (20), 
we have calculated the total number of photons in the present universe, which becomes
$$\overline N_{\gamma} = 1.74 \times 10^{88} \ . \eqno{(21)}$$
Considering the fact that the number of nucleons, $N_n$, in the
present universe is $\approx 6.30\times 10^{78}$, [11], we obtain
$$ ({\overline N_{\gamma}\over N_n}) \simeq 0.28\times 10^{10} \ . \eqno{(22)}$$
This agrees with the value $(0.14 \sim 0.33) \times 10^{10}$ as
speculated by several earlier workers [16] following calculations on baryogenesis.

To conclude, we find that the theory developed by us recently [10] for the 
evolution of the universe proves to have its further success in reproducing the 
temperature of the cosmic background radiation correctly. Besides, it also succeeds
to reproduce the photon density associated with the background radiation, and the value of
the ratio $(\overline N_{\gamma}/N_n)$, which nicely match with the results
predicted by others.
\vfill
\eject
\magnification=\magstep1
\hsize 30 pc
\vsize 42 pc
\centerline {\bf {TABLE - I }}
\midinsert$$\vbox{\offinterlineskip
\halign{&\vrule#&\strut\ #\ \cr
\noalign{\hrule}
&\hfil Age of the   \hfil&&\hfil Temperature (T) in K as  \hfil&&\hfil
Temperature (T)in K    &\hfil\cr 
&\hfil universe (t)    \hfil&&\hfil  calculated from   \hfil&&\hfil for the formation of  &\hfil\cr 
&\hfil in sec. \hfil&&\hfil  Eq. (15)   \hfil&&\hfil elementary particles    &\hfil\cr 
height3pt&\omit&&\omit&&\omit&\cr
\noalign{\hrule}
&\hfil 5   \hfil&&\hfil $\approx 1\times 10^9$   \hfil&&\hfil $\approx 6\times 10^9 (e^+,e^-)$    &\hfil\cr 
height3pt&\omit&&\omit&&\omit&\cr
&\hfil $1.2\times 10^{-4}$   \hfil&&\hfil $\approx 2.1\times 10^{11}$
\hfil&&\hfil $\approx 1.2\times 10^{12} (\mu^+,\mu^-)$ and their antiparticles   &\hfil\cr 
height3pt&\omit&&\omit&&\omit&\cr
&\hfil $7\times 10^{-5}$   \hfil&&\hfil $\approx 2.8\times 10^{11}$
\hfil&&\hfil $\approx 1.6\times 10^{12} (\pi^0,\pi^+,\pi^-)$ and
their antiparticles    &\hfil\cr 
height3pt&\omit&&\omit&&\omit&\cr
&\hfil $1.5\times 10^{-6}$   \hfil&&\hfil $\approx 1.9\times 10^{12}$
\hfil&&\hfil $\approx  10^{13}$ (protons, neutron and their antiparticles)   &\hfil\cr 
height3pt&\omit&&\omit&&\omit&\cr
&\hfil $10^{-43}$   \hfil&&\hfil $\approx 7.3\times 10^{30}$
\hfil&&\hfil $\approx  10^{32}$ (planck \ mass)   &\hfil\cr 
height3pt&\omit&&\omit&&\omit&\cr
height3pt&\omit&&\omit&&\omit&\cr
\noalign{\hrule}
\noalign{\hrule}\noalign{\smallskip}}}$$\endinsert
\vfill
\eject

\noindent {\bf References}

\item {[1]} Alan  H. Guth, in {\it Bubbles,voids and bumps in time: the new
cosmology} ed. James Cornell (Cambridge University Press,Cambridge,1989).
\item {[2]} G. Contopoulos and D. Kotsakis, {\it Cosmology},
(Springer-Verlag,Heidelberg, 1987).
\item {[3]} F. L. Zhi and L. S. Xian ,{\it Creation of the
Universe},(World Scientific,Singapur,1989).
\item {[4]} A. A. Penzias and R. W. Wilson,{\it Astrophys.Jour.} {\bf
142}, 419 (1965).
\item {[5]} D. J. Fixen, E. S. Cheng, J. M. Gales, J. C. Mather, R.
A. Shafer and E. L. Wright, Astrophysics. J. {\bf 473}, 576 (1996); 

A. R. Liddle, Contemporary Physics, {\bf 39}, no 2, 95,(1998).
\item {[6]} G. Gamow, Phys. Rev, 70, 572 (1946).
\item {[7]} J. V. Narlikar, {\it The Structure of the Universe},(Oxford
University Press, London 1977).
\item {[8]} D. N. Tripathy and Subodha Mishra, Int. J. Mod. Phys. D {\bf
7}, 3 , 431 (1998).
\item {[9]} S. Chandrasekhar, {\it Monthly Notices Roy.Astron. Sco}. {\bf 91},456 (1931).
\item {[10]} D. N. Tripathy and Subodha Mishra, Int. J. Mod. Phys. D {\bf 7}, 6, 917 (1998).
\item {[11]} P. S. Wesson, {\it Cosmology and Geophysics},(Adam Hilger Ltd,Bristol,1978).
\item {[12]} E. R. Harrison,{\it Cosmology}, (Cambridge: Cambridge University Press,1981).
\item {[13]} D. A. McQuarrie, {\it Statistical Mechanics}, (Harper
 \& Row, New York, 1976).
\item {[14]} R. K. Pathria, {\it Statistical Mechanics},(Pergamon
Press, Oxford, 1972).
\item {[15]} A. M. Boesgaard and G. Steigman, Ann. Rev. Astron. {\bf
23}, 319 (1985).
\item {[16]} I. Affleck and M. Dine, Nucl. Phys. {\bf B249}, 361 (1985).

\vfill
\eject
\end